\documentclass[%
 reprint,
superscriptaddress,
 amsmath,amssymb,
 aps,
pre,
]{revtex4-2}

\usepackage{epsfig,dsfont,amssymb,amsmath,amsthm,amsfonts,amsbsy,mathrsfs}
\usepackage{graphicx}
\usepackage{color}
\usepackage{bm}
\usepackage{multirow}
\usepackage{natbib}

\newcommand{\rr}{{\mathbf r}}

\newcommand{\BEQ}{\begin{equation}}
\newcommand{\EEQ}{\end{equation}}
\newcommand{\BEA}{\begin{eqnarray}}
\newcommand{\EEA}{\end{eqnarray}}

\newcommand{\qv}{\mathbf{q}}

\usepackage{tikz}
\usetikzlibrary{arrows,shapes,chains}

\begin{document}
\preprint{APS/123-QED}

\title{From Motility-Induced Phase-Separation to Glassiness in Dense Active Matter}

\author{Matteo Paoluzzi}
\email{matteopaoluzzi@ub.edu}
\affiliation{Departament de Física de la Mat\`eria Condensada, Universitat de Barcelona, C. Martí Franqu\`es 1, 08028 Barcelona, Spain.}

\author{Demian Levis}

\affiliation{Departament de Física de la Mat\`eria Condensada, Universitat de Barcelona, C. Martí Franqu\`es 1, 08028 Barcelona, Spain.}
\affiliation{UBICS University of Barcelona Institute of Complex Systems, Mart\'i i Franqu\`es 1, E08028 Barcelona, Spain.}

\author{Ignacio Pagonabarraga}
\affiliation{CECAM Centre  Europ\'een  de Calcul Atomique et Mol\'eculaire, \'Ecole Polytechnique F\'ed\'erale de Lausanne (EPFL),
Batochime, Avenue Forel 2, 1015 Lausanne, Switzerland.}
\affiliation{Departament de Física de la Mat\`eria Condensada, Universitat de Barcelona, C. Martí Franqu\`es 1, 08028 Barcelona, Spain.}
\affiliation{UBICS University of Barcelona Institute of Complex Systems, Mart\'i i Franqu\`es 1, E08028 Barcelona, Spain.}

\date{\today}

\begin{abstract}
Dense active systems are widespread in nature, examples range from bacterial colonies to biological tissues. Dense clusters of active particles can be obtained by increasing the packing fraction of the system or taking advantage of a peculiar phenomenon named motility-induced phase separation (MIPS). 
In this work, we explore the phase diagram of a two-dimensional model of active glass and show that disordered active materials develop a rich collective behaviour encompassing both MIPS and glassiness.
We find that, although the glassy state is almost indistinguishable from that of equilibrium glasses, the mechanisms leading to its fluidization do not have any equilibrium counterpart. 
Our results can be rationalized in terms of a crossover between a low-activity regime, where glassy dynamics is controlled by an effective temperature, and a high-activity regime, which drives the system towards MIPS. 
\end{abstract}

\maketitle

\section*{Introduction}
Active self-propelled particles have been often employed as a starting and minimal model
for capturing collective behaviors in biological and living materials, examples run from biological tissues to dense drops of ants \cite{klopper2018physics,trepat2018mesoscale,feinerman2018physics,Marchetti13,Bechinger17}.  Moreover, active agents might show a high degree of heterogeneities in their microscopic characteristics inside a given population.  However, in modeling biological tissues, usually, it is assumed that each cell has the same mechanical and geometrical properties \cite{trepat2018mesoscale}. Cellular differences play a crucial role in many biological processes as in the case of cancer development, where cell heterogeneity is a common feature \cite{marusyk2010tumor}. Phenotypic heterogeneity has functional consequences that impact the ability of microbes to adapt to different environments \cite{ackermann2015functional} and mechanical heterogeneity changes the collective properties in simple models of biological tissues \cite{PhysRevLett.123.058101}.

Living systems composed of a collection of self-propelled agents are thus naturally characterized by a certain degree of variability in size, shape, mechanical properties, and motility parameters. Often, in the system of interest, only average values are available: it is thus  reasonable to introduce suitable probability distributions for those quantities. 

In this work, we will focus our attention on the effect of quenched fluctuations in the typical active agent size. To take into account this aspect, we perform numerical simulations of Active Brownian disks of heterogeneous size. To make a contact with what is known on equilibrium systems, we study the phase diagram of repulsive Active Brownian Particles (ABP) composed of a continuous, i. e., polydisperse, mixture \cite{PRX_Berthier}. Systems of polydisperse self-propelled particles play an important role for our understanding of dense and active materials \cite{Henkes2011,szamel2015glassy,flenner2016nonequilibrium,berthier2019glassy, janssen2019active, KumarPoly2021,preprint}. On the other hand, understanding the glass transition in active matter requires a huge theoretical and numerical effort and many progresses in the framework of Mode-Coupling Theories have been done during the last years \cite{fily2014freezing,reichert2021transport,de2018static,PhysRevE.104.044608,reichert2021tracer}.

 The model system introduced here is particularly suitable for studying the interplay between Motility-Induced Phase Separation (MIPS) and glassy dynamics \cite{cates2015motility,ni2013pushing,berthier2014nonequilibrium,PhysRevLett.125.218001,nandi2018random,mandal2020extreme}. 
MIPS is a well-established phenomenon in Active Matter whose main features are very robust against different microscopic active dynamics \cite{PhysRevLett.121.098003,maggi2021universality} and dimensionality of the space \cite{PhysRevLett.126.188002}.
We shall see that glassiness occurs only at low activity, although the system in the MIPS phase arranges in dense patches that do not provide any evidence of positional order (and thus glassy behavior might develop).

Using as a control parameter the persistence length {$\ell$} and the packing fraction $\phi$, we study the phase diagram of polydisperse ABP and investigate the properties of the system at high packing fractions, where we document a glass transition driven by the persistence length. The latter is signaled by a dynamical slowing down of the time correlation function of the hexatic order parameter, with a decoupling between density and hexatic fluctuations. We show that, in the glassy region, there is a coexistence of hexatic and liquid phases. We then look at the statistical properties of the inherent structures obtained  from stationary configurations at different parent activities. 
In this way, we can provide an estimate of the distance, computed in a probabilistic sense, between a typical steady-state configuration, that results from the non-equilibrium dynamics in the presence of self-propulsion, and the corresponding configuration that minimizes the mechanical energy.
We show that glassy configurations are almost inherent configurations. Comparing dense configurations obtained for higher persistence length, we document a discontinuous crossover approaching MIPS.

Finally, 
we explore the concept of effective temperature in connection with the vibrational density of states of instantaneous configurations. 
In general, the eigenvalues of the dynamical matrix can be either positive or negative in the case of
instantaneous configurations. Since negative eigenvalues are connected to unstable directions in the instantaneous energy landscape, they can provide information on barrier crossing \cite{keyes1997instantaneous,stratt1995instantaneous,bembenek1996role}.

We show that the magnitude of the largest negative eigenvalues of the dynamical matrix, i. e., the {\it "faster"} unstable mode of the system, encodes information about the break down of the concept of effective temperature in active glasses. Those unstable directions can be surfed by the active system for removing geometrical frustration and bringing the system into a fluid phase.
This result suggests that, even though glassy configurations at low activity are basically indistinguishable by their equilibrium counterpart, the mechanism of fluidization in active systems might be distinct from that in equilibrium glasses, where mean-field models point the attention out on marginal directions in the potential energy landscape rather than the unstable ones \cite{castellani2005spin}. 

In the present work, as main results we obtain that (i) the MIPS region extends up to very large packing fractions, (ii) there is no hint of dynamical slowing down in the MIPS regime, even if the dense MIPS phase is an assembly of polydisperse particles at high packing fraction, (iii) the mechanisms of fluidization of the active systems might be quantitatively connected with unstable modes of the instantaneous spectrum.

\section*{Results}
\subsection*{Absence of structural arrest at high activity}
\begin{figure*}[!t]
\centering\includegraphics[width=.45\textwidth]{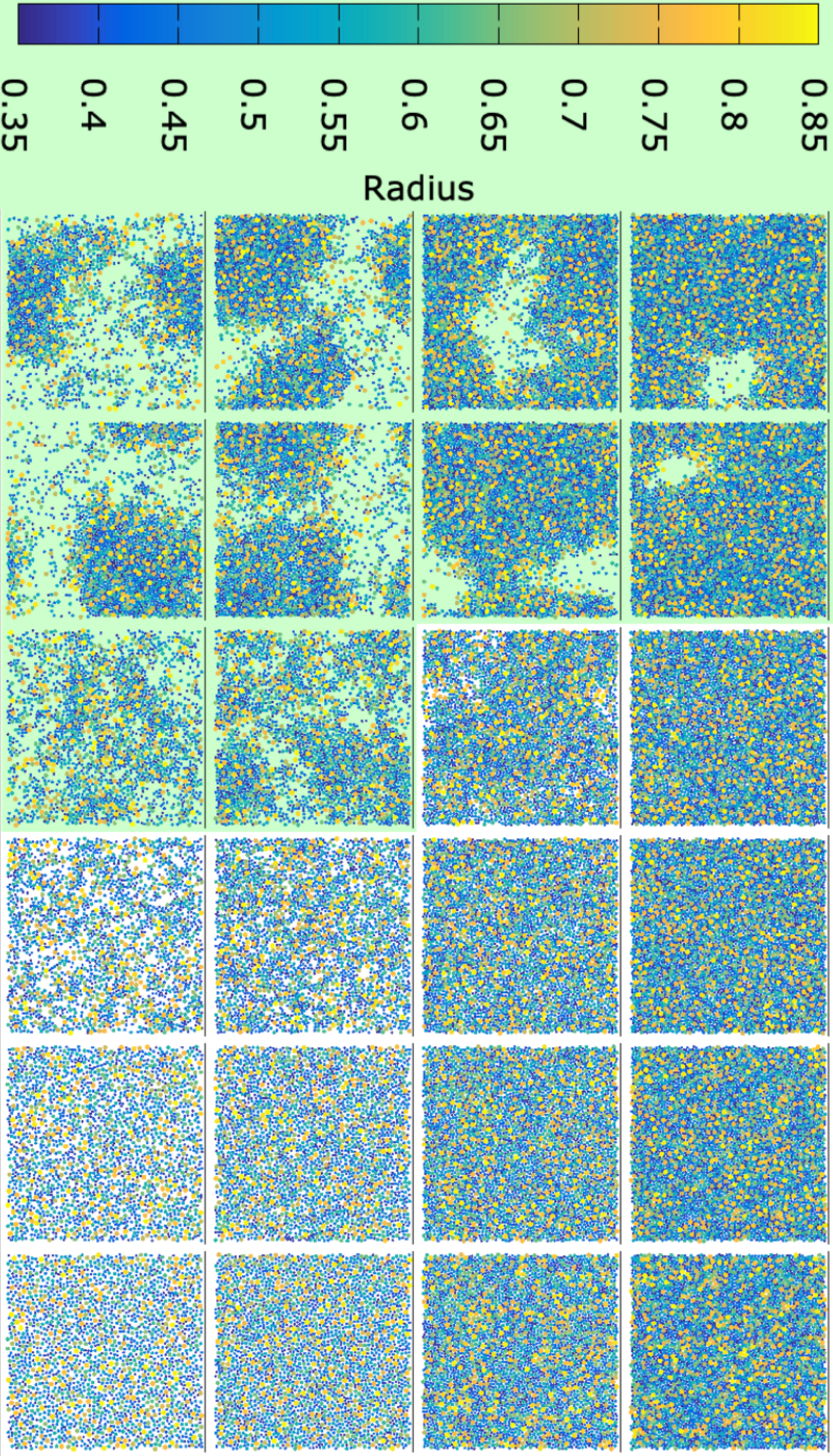}
\caption{ Representative stationary configurations. The packing fraction increases from left to right $\phi \simeq 0.4,0.5,0.7,0.8$. The persistence length increases from bottom to top $\ell=0.01,0.1,1,10,50,200$. Particle colors indicate their radius (increasing values from violet to yellow).
For large persistence lengths, the system undergoes Motility-Induced phase separation (MIPS) (snapshots with green background). The MIPS region extends to large packing fractions, i. e., $\phi \simeq 0.8$. In the dense regime, the system behaves as a glass at small persistence lengths, then melts to a fluid state, and finally undergoes MIPS.}
\label{fig:snap_pd}      
\end{figure*}

We start our study  considering a system composed of fully polydisperse Active Brownian Particles (ABPs, details about the microscopic model are provided in Materials and Methods).
A qualitative view of the phase diagram is provided in Fig. \ref{fig:snap_pd} where we report the snapshots of typical stationary configurations. As one can see, the system undergoes MIPS and separates into a dense and dilute phase for large enough values of the persistence length $\ell$. We observe that the MIPS regime extends until the largest packing fraction simulated, i. e., $\phi = 0.78$. As we shall study in the next sections, the high density regime is characterized by a glassy phase at small persistence lengths.

\begin{figure*}[!t]
\centering\includegraphics[width=1.\textwidth]{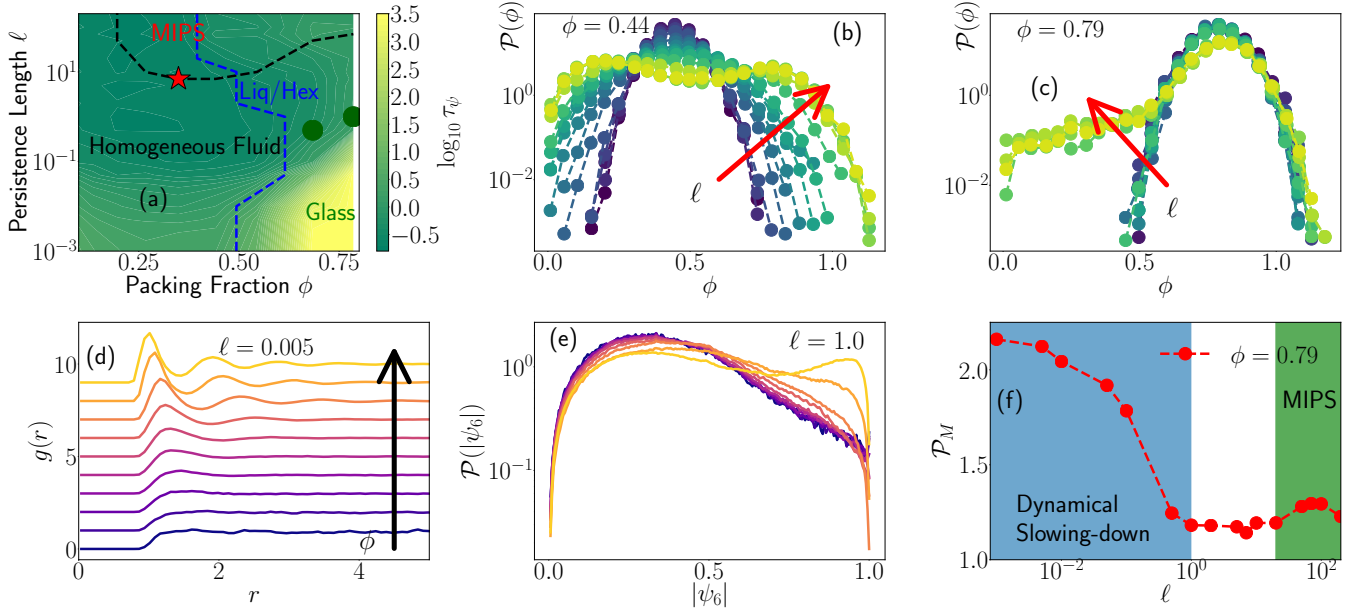}
\caption{ Phase diagram of fully polydisperse Active Brownian particles. (a) The black dashed curve individuates the Motility-Induced Phase Separation
(MIPS) coexistence region. The red star indicates the critical point.  The blue dashed curve corresponds to the transition between homogeneous fluid and liquid/hexatic coexistence. 
The color map indicates the relaxation time (in log scale). Green symbols indicate the typical persistent length corresponding to the smaller negative eigenvalue of the energy landscape.
(b) Probability distribution function of the local packing fraction $\mathcal{P}(\phi)$ crossing the MIPS region ($\phi=0.44$). The red arrows indicate increasing values of $\ell=0.001,0.005,0.01,0.05,0.1,0.5,1,2,5,7,10,20,50,70,100,200$, from violet to yellow, respectively. (c) $\mathcal{P}(\phi)$ at high densities ($\phi=0.79$),  where the system undergoes a dynamical slowing down at low activities (the glass region in (a)). Colors and persistence length values same as in (b). 
 (d) Radial distribution function$g(r)$ for small persistence length ($\ell=0.005$) as density increases (as indicated by the black arrow, $\phi=0.05, 0.09, 0.14, 0.20, 0.27, 0.35, 0.44, 0.54, 0.68, 0.79$). For  clarity,  curves  have  been  shifted  vertically by $1\times n$, with $n=0,1,...,9$ labelling each curve.
 Approaching the glass transition, $g(r)$ does not reveal the presence of any solid-like positional order, as in the case of a fluid. 
(e) Probability distribution function of the hexatic order parameter $\mathcal{P}(|\psi_6|)$ at intermediate persistence length ($\ell=1.0$) for increasing values of density, (increasing values of density $\phi=0.05, 0.09, 0.14, 0.20, 0.27, 0.35, 0.44, 0.54, 0.68, 0.79$ from violet to yellow as in (d)). 
Crossing the blue dashed line in panel (a), the distribution develops a double-peaked structure. (f) Height of the high-$|\psi_6|$ peak of $\mathcal{P}(|\psi_6|)$ as persistence length increases for $\phi=0.79$. The non-monotonous behavior (the peak decreases approaching a minimum value in the homogeneous phase and thus increases again in the MIPS region) suggests
the presence of a re-entrance in the phase diagram (error bar smaller than the symbols, average over stationary trajectories and $N_s=10$ independent runs are considered). 
}
\label{fig:pd}      
\end{figure*}

We start our quantitative analysis by looking at the statistical properties of the relevant coarse-grained fields, i. e., the local packing fraction field $\phi(x,y,t)$, which measures the packing fraction around the point of coordinate $(x,y)$ at  time $t$, and  hexatic order parameter $\psi_6(x,y,t)$, which provides a local measure of the orientational order in $(x,y)$ at time $t$. Details about their definition and their computation are provided in the section Methods. For monitoring the presence of MIPS, we look at the behavior of the probability distribution function of $\phi(x,y,t)$,  $\mathcal{P}(\phi) = \langle \delta(\phi - \phi(x,y,t)) \rangle_t $ (see Materials and Methods for details). To investigate the emergence of a disordered glassy phase, we study the relaxation time of density fluctuations through the intermediate scattering function $F_{self}(q,t)$ and the time-correlation of the hexatic order parameter $C_\psi(t)$ (their definition is provided in Methods).  
The resulting phase diagram  is shown in Fig. \ref{fig:pd}a where the color map indicates the structural relaxation time measured through $C_\psi$ (using $\tau_\alpha$ defined from the decay of $F_{self}$ does not introduce any qualitative changing in the phase diagram). As one can appreciate, although in the MIPS region the system separates into a dilute and a dense phase (with a dense phase reaching packing very high packing fractions, as shown by the behavior of $\mathcal{P}(\phi)$), the glassy region is limited to high packing fraction and small persistence length. It is worth noting that, in systems composed of Active-Ornstein Uhlenbeck particles, where the natural control parameters are the correlation time of the noise and its strength \cite{Paoluzzi}, keeping constant the diffusion constant, one can obtain a glassy regime that goes to very large persistence lengths \cite{preprint}.

The behavior of 
$\mathcal{P}(\phi)$ crossing the MIPS region is shown Fig. \ref{fig:pd}b. The distribution develops the typical double-peaked structure due to the presence of two coexisting phases. Since the potential is very steep and soft, the tail of the peak at high packing fractions can reach  $\phi \gtrsim 1$. The same is true even at very high densities, as documented in Fig. \ref{fig:pd}c.
The qualitative features of the phase diagram are consistent with those observed in athermal self-propelled disks \cite{fily2014freezing}.
We checked that the system remains always in disordered, liquid-like configurations, in the whole range of parameters, by looking at the radial distribution function $g(r)$, as shown in Fig. \ref{fig:pd}d.
$g(r)$ does not reveal any hint of crystallization, even at the largest packing fraction $\phi=0.78$.

Polydisperse disks introduce geometrical frustration in the system, inhibiting crystallization. Although at high density and small persistence length a glassy regime develops, heterogeneous regions where the system develops hexatic patterns  survive over short length scales (of the order of a few particle radii) at high densities. The presence of these heterogeneous zones of high and low $\psi_6$ values determine a region of the phase diagram that is delimited by the dashed blue line in Fig. 2a, and that we identify with the label Liquid/Hexatic. 
To investigating this feature, we explore the presence of hexatic patches, typical of two-dimensional systems \cite{PhysRevLett.121.098003, kawasaki2007, flenner2015fundamental, PhysRevLett.125.178004}, by looking at the distribution $\mathcal{P}(|\psi_6|) = \langle \delta(|\psi_6| - |\psi_6(x,y,t)|) \rangle_t$, shown in Fig. \ref{fig:pd}e ($\ell=0.1$). At high densities, the distribution becomes double peaked indicating the presence of hexatic patterns.

Crossing the dashed blue line in Fig. 2a, $\mathcal{P}(|\psi_6)|$ changes from single to double-peaked.
The Fluid/Hexatic transition seems to develop a reentrance in the phase diagram, as suggested by the behavior of the orientational order parameter $\langle |\psi_6|\rangle $ all along the phase diagram (see Supplementary Note 1, where the contour plot of $\langle |\psi_6|\rangle$ is shown). This is highlighted in Fig. \ref{fig:pd}f where we report the height of the second peak, i. e., $\mathcal{P}_M \equiv max_{|\psi_6|>0.5}\mathcal{P}(|\psi_6|)$, as a function of $\ell$ at $\phi=0.79$. Within the statistics considered here, $\mathcal{P}_M$ is a non-monotonous function of $\ell$, i. e., $\mathcal{P}_M$ decreases as $\ell$ increases towards the fluid phase and, eventually, develops a bump in the MIPS region. This behavior suggests that
hexatic patches might be more pronounced in the glassy and MIPS region 
than in the liquid region.

\subsection*{Stability of MIPS against quenched fluctuations}
In the previous section we have introduced the phase diagram of the fully polydisperse model. Now we are going to investigate the impact of geometrical frustration on the MIPS dense phase.
In order to quantify the effect of particle heterogeneities on MIPS,
we have performed numerical simulations deep in the phase-separated region for a larger system size, i. e., $\ell=100, \, \phi=0.44, \, L=120\langle \sigma \rangle$, and by varying the fraction $f\in[0,1]$ of polydisperse disks. This means that at $f=0$, the system is monodisperse, i. e., $\sigma_i = \langle \sigma \rangle$, $\forall i$.

\begin{figure*}[!t]
\centering\includegraphics[width=1.\textwidth]{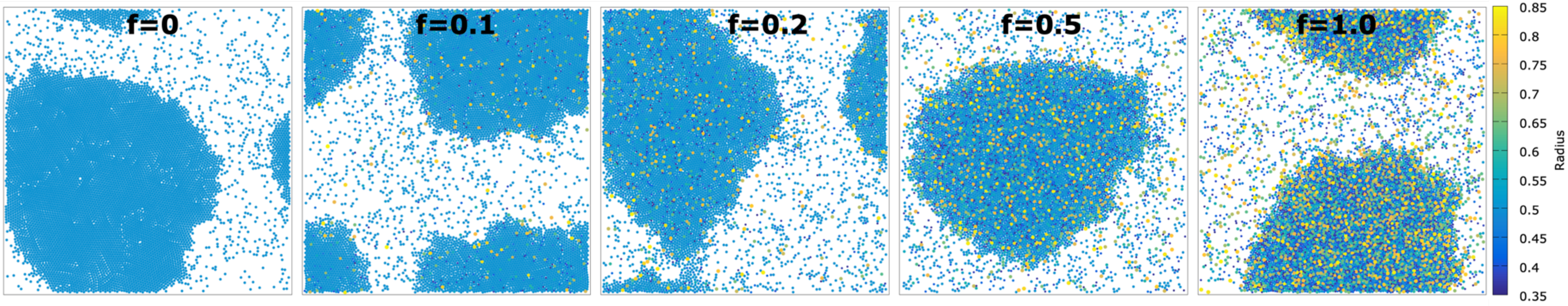}
\caption{Representative stationary configurations in Motility-Induced Phase Separation.
Snapshots of mixture of Active Brownian particles composed by a fraction $f$ of polydisperse disks ($\ell=100$, $\phi=0.44$, $L=120\langle \sigma \rangle$). Particle colors indicate their radius.}
\label{fig:mips_mix_snap}      
\end{figure*}

\begin{figure*}[!t]
\centering\includegraphics[width=1.\textwidth]{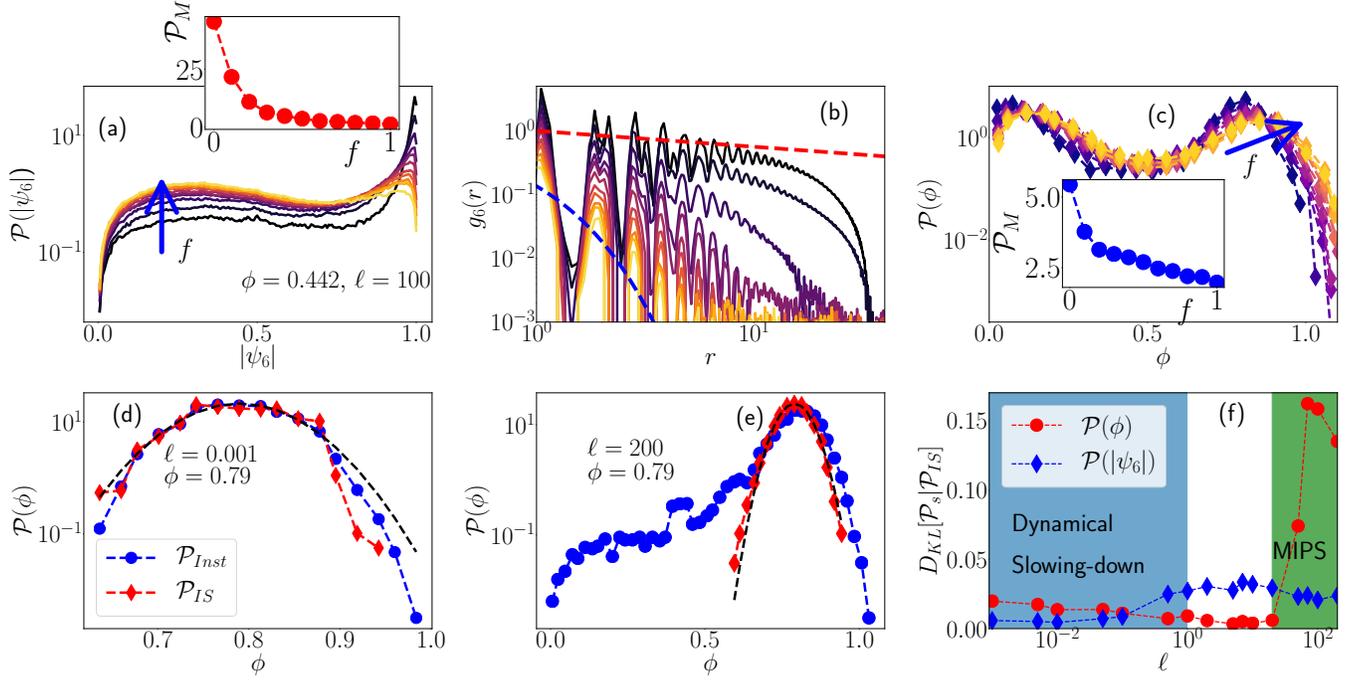}
\caption{The impact of geometrical heterogeneities on Motility-Induced Phase Separation. 
(a) Probability distribution of the hexatic order parameter $\mathcal{P}(|\psi_6|)$ by varying the fraction of heterogeneous particles from $f=0$, to $f=1$, from black to orange (the arrow points increasing values of $f=0.0,0.1,0.2,...,1.0$). The inset shows the height of the second peak $\mathcal{P}_M\equiv $ (at high $|\psi_6|$ values) as a function of $f$. 
(b) Spatial correlation function $g_6(r)$ as $f$ increases, from black to orange. The dashed red curve is a power law $r^{-1/4}$, the dashed blue curve is an exponential decay $e^{-r/\xi}$ with $\xi \simeq O(1)$.
(c) Probability distribution of the local packing fraction $\mathcal{P}(\phi)$ by varying $f$ (the arrow points increasing values of $f$). The inset shows the height of the peak at high $\phi$ values as $f$ increases. (d) $\mathcal{P}(\phi)$ for small persistence length ($\ell=0.001$) and high density ($\phi=0.79$), blue circles refer to instantaneous configurations, red diamonds to inherent configurations. The dashed black line is the fit to a Gaussian function. (e) 
$\mathcal{P}(\phi)$ of instantaneous and inherent configurations for large persistence length ($\ell=200$ and $\phi=0.79$). $\mathcal{P}_{IS}(\phi)$ is well captured by a Gaussian fit (dashed black line). (f) Kullback–Leibler divergence $D_{KL}$ between stead-state ($\mathcal{P}_S$) and inherent ($\mathcal{P}_{IS}$) distributions for the two observables, i. e., $\mathcal{P}(\phi)$ (red circles) and $\mathcal{P}(|\psi_6|)$ (blue diamonds).}
\label{fig:mips_mix}      
\end{figure*}

Typical stationary configurations are shown in Fig. \ref{fig:mips_mix_snap}.
A quantitative analysis shows that, although $\mathcal{P}(|\psi_6|)$ remains double-peaked at any $f$ value, a small fraction of heterogeneous disks causes a huge decrease of the height of the second peak (see Fig. \ref{fig:mips_mix}a). 
This fact can be quantified further by looking at $\mathcal{P}_M$ defined before.  As one can see in the inset of Fig. \ref{fig:mips_mix}a, $\mathcal{P}_M$ dramatically decreases as $f$ increases. The reduction of orientational order can be appreciated by the study of $g_6(r)$.
The typical behavior of $g_6(r)$ as $f$ increases from $0$ to $1$ is presented in  Fig. \ref{fig:mips_mix}b. Although, because of finite size effects, it is hard to identify a clear power-law behavior at $f=0$ \cite{wilding}.
As $f$ increases, $g_6(r)$ tends to decay exponentially, i. e., $g_6(r) \sim e^{-r/\xi}$, with a correlation length of the order of a few particle sizes that tends to screen any power-law decay. 
The fact that the system loses orientational order on large scales does not ensure the absence of a dynamical arrested phase. The latter is excluded by the behavior of the relaxation time of $C_\psi$ (the color map in the phase diagram for $f=1$ shown in Fig. 2a) that provides clear evidence of a liquid-like dense phase (excluding any kind of glassy dynamics in the dense phase of MIPS).
However, such structural change in the hexatic patches is not accompanied by any dramatic change on the density de-mixing due to MIPS. 
This is shown in Fig. \ref{fig:mips_mix}c, where we report $\mathcal{P}(\phi)$ for different values of $f$. We observe that height of the high-density peak decreases as the fraction of heterogeneous disks increases. However, the position of the two peaks remains much more stable, indicating that both, the dense and the dilute phase, remain at almost the same average densities, i.e. the MIPS coexistence region remains largely unaffected by polydispersity. This effect can be quantified looking at height of the high-density peak of $\mathcal{P}(\phi)$, as it is shown in the inset of Fig. \ref{fig:mips_mix}c. Although the peak slightly reduces its height as $f$ increases, the distribution develops a fat tail for high $\phi$ values indicating a robust phase separation in dense and dilute regions.

For gaining insight into the nature of the dense phase, we have computed the inherent structures at different parent activities. In practice, we consider an instantaneous configuration taken in the stationary state at $\phi=0.79$ for a given value of $\ell$.
We then compute the corresponding inherent configuration by minimizing the configurational energy $\Phi$ that is solely due to the pair interactions. 

We start our study comparing the structural properties of inherent and instantaneous configurations.  
In order to perform a quantitative analysis, we compared the probability distribution functions $\mathcal{P}(\phi)$ and $\mathcal{P}(|\psi_6|)$ resulting from the inherent structures, i. e., $\mathcal{P}_{IS}$, with those obtained from instantaneous configurations, i. e., $\mathcal{P}_{inst}$. 

The behavior of $\mathcal{P}_{IS,inst}(\phi)$ is shown in Fig. \ref{fig:mips_mix}d for $\ell=0.001$. In Fig. \ref{fig:mips_mix}e the same observables are evaluated for $\ell=200$. As one can see, inherent and instantaneous configurations are quite similar at low persistence length ($\ell=0.001$), proving that the system reached a mechanically stable configuration that is slightly perturbed  by activity. 
Different is the situation at high parent activities, i. e., $\ell=200$, Fig. \ref{fig:mips_mix}e, where instantaneous configurations taken in the stationary state reveal the presence of MIPS through a broad tail in $\mathcal{P}_{inst}(\phi)$. 
On the other hand, the inherent configurations are homoheneous and thus they produce a Gaussian distribution centered around $\phi=0.79$.

For making quantitative progresses we measure the distance between $\mathcal{P}_{IS}$ and $\mathcal{P}_{ins}$ using
the Kullback–Leibler divergence $D_{KL}[\mathcal{P}_{IS} | \mathcal{P}_{inst}]$ (see Methods). The result is shown in Fig. \ref{fig:mips_mix}f for both  $\varphi$ and $\psi_6$. We obtain that $D_{KL}[\mathcal{P}_{IS}(\varphi) | \mathcal{P}_{inst}(\varphi)]$ is small in the glassy state, at small $\ell$, indicating that instantaneous and inherent configurations  are almost identical, and even decreases as  $\ell$ increases. However, as the system approaches MIPS, $D_{KL}$ jumps to higher values. This is because inherent configurations are homogeneous while instantaneous ones tend to decompose into two phases because due to MIPS. Less informative is $D_{KL}[\mathcal{P}_{IS}(|\psi_6|) | \mathcal{P}_{inst}(|\psi_6|)]$ that maintains  small values showing a smooth crossover on intermediate persistence length, i. e., $\ell \sim 0.1$. This crossover reveals measurable differences in the orientational order between inherent and instantaneous configurations.

\subsection*{Activity-driven dynamical arrest}
\begin{figure*}[!t]
\centering\includegraphics[width=1.\textwidth]{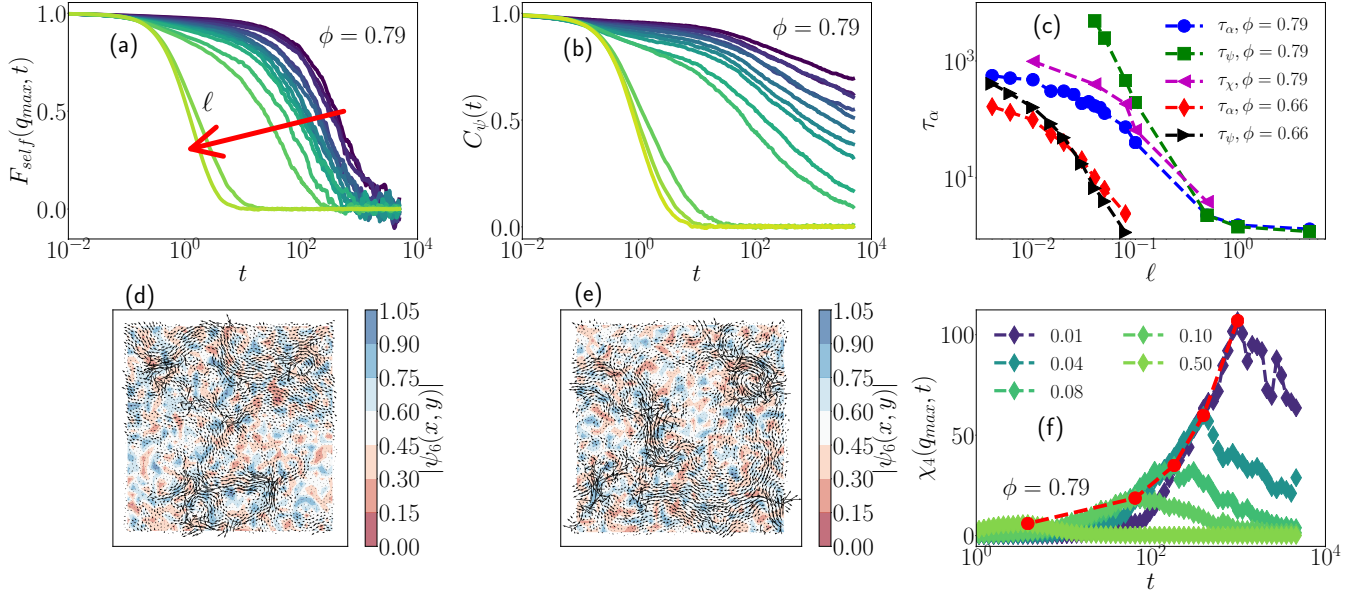}
\caption{ Relaxation dynamics at high packing fraction fully polydisperse Active Brownian particles. (a) Intermediate scattering function $F_{self}(q,t)$ computed at the peak of the static structure factor $q=q_{max}$. 
The red arrow indicates increasing values of $\ell=0.004,0.006,0.01,0.015,0.02,0.025,0.03,0.035,0.04,0.045,0.05,0.08,0.1,0.5,1.0,5.0$
(b) Relaxation dynamics of the correlation $C_\psi(t)$. (c) Structural relaxation time obtained from the intermediate scattering function ($\tau_\alpha$), from $C_\psi$ ($\tau_\psi$), and from the peak of the dynamical susceptibility ($\tau_\chi$). Density fluctuations decay faster than the orientational ones. The map of displacements on a time scale $\tau_{\chi}$ are shown in (d), and (e), for $\ell=10^{-3}$ and $10^{-2}$, respectively. The color map indicates the magnitude of the hexatic order field $|\psi_6(x,y)|$. The map reveals the presence of heterogeneous regions in both fields, long-time displacement and local hexatic order.
(f) Dynamical susceptibility $\chi_4(t)$ for different persistent lengths $\ell$ (see legend). The red circles indicate the peak position $\chi_4(\tau_\chi)$ ($\tau_\chi$ is reported in (c) as a function of $\ell$).
}
\label{fig:glassy}      
\end{figure*}

We now explore the high packing fraction region $\phi=0.79$ in the case of fully polydispersisity, i. e., $f=1$. 
We study the behavior of the self-part of the intermediate scattering function $F_{self}(q,t)$, and the time-correlation function of the hexatic order parameter $C_\psi(t)$. 
We monitor $F_{self}(q_{max},t)$, where $q_{max}$ is the wave vector of the first peak of the static structure factor,  for increasing values of $\ell$, see Fig. \ref{fig:glassy}a. 
As one can see, activity tends to melt the system. 
However, as well as in the case of other two dimensional glassy systems \cite{flenner2015fundamental}, $F_{self}(q,t)$ does not show a clear two-step decay, indicating that the cage effect is not as strong as in the case of a three-dimensional system. Looking at the sample-to-sample fluctuations of $F_{self}(q,t)$, we obtain a dynamical susceptibility $\chi_4(q,t)$ that behaves as in the case of an ordinary supercooled liquid (Fig. \ref{fig:glassy}f) \cite{Berthier2011}. The broad peak in $\chi_4(q,t)$ reveals the presence of dynamical heterogeneity, as shown Fig. \ref{fig:glassy}d and Fig. \ref{fig:glassy}e,
for $\ell=0.001,0.01$, respectively, where we 
show the map of displacements, i. e., each arrow indicates the typical displacement performed by a particle on the time scale $\tau_\chi$, with $\tau_\chi$ obtained from the peak of $\chi_4(q,t)$.
It is worth noting that $\tau_\chi$ behaves quantitatively in the same way as $\tau_\alpha$, i. e., the structural relaxation time obtained from $F_{self}$, (see Fig. \ref{fig:glassy}c). 

A more clear two-step decay is visible in $C_\psi(t)$ (Fig. \ref{fig:glassy}b). This fact suggests the slowest structural degree of freedom of the system is the hexatic order parameter rather than the density. This claim is supported by the evidence that there is a bifurcation of the relaxation time scales of the two observables approaching the glass transition (Fig. \ref{fig:glassy}c).  
We notice that similar differences in relaxation of $F_{self}$ and $C_\psi$ are documented in two-dimensional equilibrium glasses \cite{vivek2017long,illing2017mermin} where Mermin-Wagner excitations play an important role \cite{PhysRevLett.117.245701}.
Because of the decoupling between $\tau_\chi$ and $\tau_\psi$, the order parameter $\psi_6$ remains frozen on the typical time scale of dynamical heterogeneities, as highlighted by the color map in (d) and (e). 
It is worth noting that, as it has been shown in Fig. \ref{fig:pd}f, 
the height of the peak at higher $
|\psi_6|$ values tend to reduce is height as the persistence length $\ell$ increases reaching a minimum at $\ell \sim 1$. As we will discuss later, this characteristic value of $\ell$ is related with the properties of the instantaneous normal modes.
The same scenario has been observed at lower densities, i. e., $\phi = 0.66$.

\subsection*{Instantaneous normal modes and the breaking of the effective temperature}
The concept of effective temperature helps to rationalize 
some features of active systems \cite{szamel2014self,Maggi14,PhysRevE.84.040301,levis2015,maggi2017memory,henkes2020dense,Bi2016,giavazzi2018flocking,petrelli2020}. 
In colloidal glasses driven by thermal noise, because of the caging effect, particles spend most of the time vibrating around their equilibrium position until cooperative rearrangements allow the particle to escape from the cage \cite{leuzzi2007thermodynamics}.
This equilibrium-like picture can be reasonably employed also in the case of an active glass whenever the active motion causes only vibrations and thus rattling inside the cage, i. e., in the small persistence length regime $\ell \leq \langle \sigma \rangle$.

In order to make progress, we consider harmonic vibrations around an equilibrium inherent configuration. The harmonic vibrations define a set of eigenfrequencies $\omega_\nu^2=\lambda_\nu$, with $\lambda_\nu$ the the $\nu-$th eigenvalue of the dynamical matrix \cite{henkes2020dense,PhysRevE.84.040301,Bi2016}. We can thus introduce the following (mode-dependent) effective temperature \cite{henkes2020dense} (see Supplementary Note 2)  \cite{Paoluzzi,PaoluzziXY,PhysRevResearch.2.023207}) $T_{{eff}}(\lambda_\nu,\ell) =\frac{v_0^2}{2(v_0 / \ell + \mu \lambda_\nu )}    \;$, with $\mu$ being the particles' mobility.
At variance with early studies, here we are going to provide an alternative interpretation of $T_{{eff}}(\lambda_\nu,\ell)$ that allows to gain insight into the active glass transition.

Let $\rr^{*}\equiv (\rr_1^*,\rr_2^*,...,\rr_N^*)$ 
be the inherent configuration of the system so that $\left. \partial_i \Phi \right|_{\rr^{*}}=0$. The stability of such a configuration is written in the eigenvalues $\lambda_\nu$ of the Hessian matrix $\mathbb{H}_{ij} = \partial_i \partial_j \Phi$. If all the eigenvalues $\lambda_\nu$ are strictly positive, the configuration lays in a minimum of the potential energy landscape.
{Let us introduce the persistent time $\tau=\ell /v_0 $.}
As soon as the system develops instabilities, i. e., directions with negative curvature in the energy landscape, there will be a critical value $\tau_c$ that causes a divergence in $T_{eff}$ (for a given negative value $\lambda_\nu$, we can write $\lambda_\nu=-|\lambda_\nu|$ and $T_{eff}$ loses completely sense for $\tau - \mu |\lambda_\nu|=0$) so that, looking at instantaneous configurations that naturally develop unstable directions, $T_{eff}$ becomes ill-defined for increasing values of $\tau$.
This critical value corresponds to the largest negative eigenvalue $\lambda_{max}^{(-)}$, i. e., $\tau_c = |\lambda_{max}^{(-)}|^{-1}$. For typical configurations in the liquid regime, the energy spectrum always develops negative eigenvalues, making the concept of effective temperature meaningful only in the limit $\tau \ll \tau_c$.

\begin{figure*}[!t]
\centering\includegraphics[width=.45\textwidth]{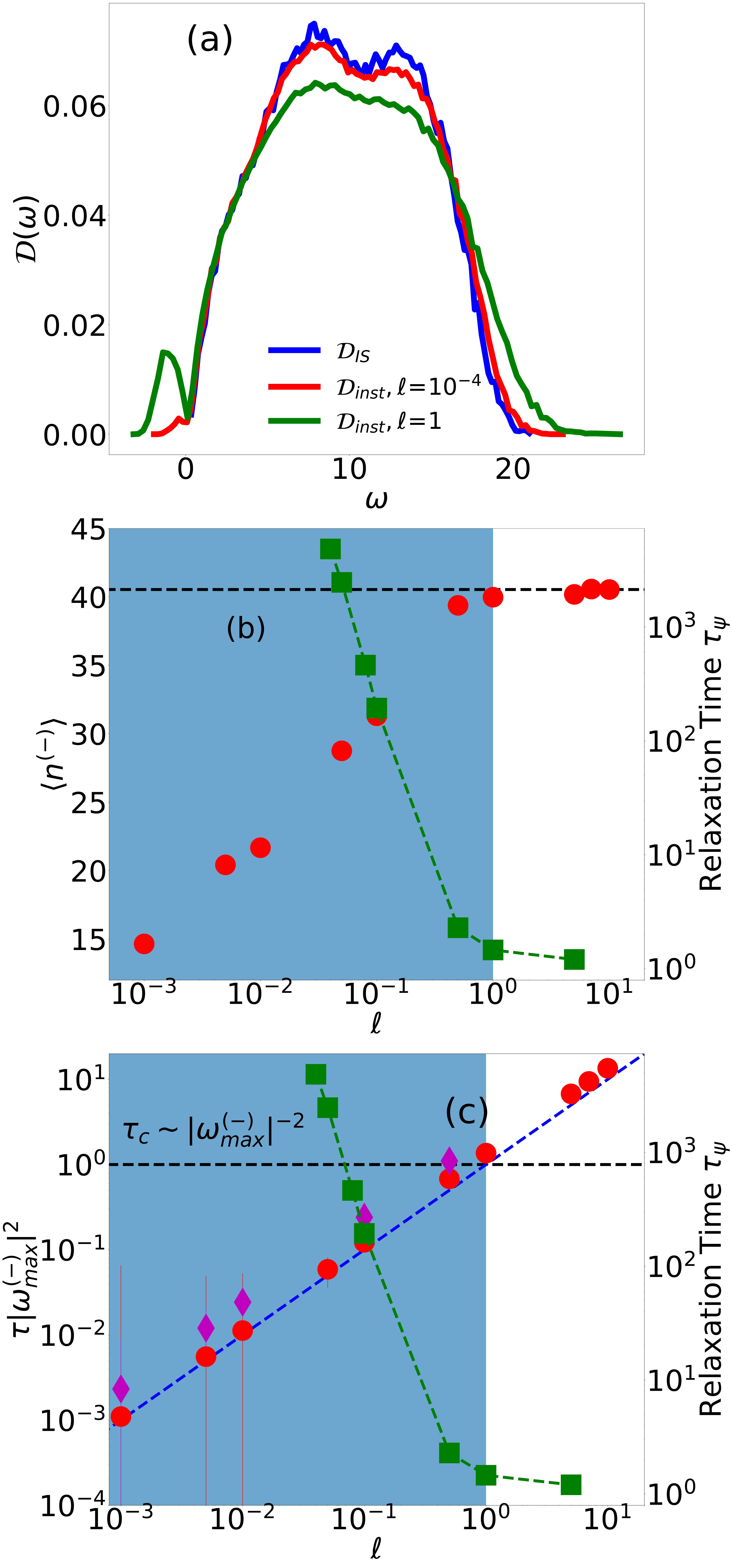}
\caption{Instantaneous Normal Modes of fully polydisperse Active Brownian particles. (a) $\mathcal{D}(\omega)$ of inherent configurations (blue curve) and instantaneous configurations at different persistence length ($\ell=10^{-4},1$, red and green, respectively.) (b) The average number of negative frequencies (red circles, error bar smaller than symbols, average over $100$ independent configurations) approaches a plateau value at large $\ell$ and undergoes a crossover for $\ell\sim 1$ at the same point where the structural relaxation time (green squares) starts to grow for decreasing values of $\ell$ (blue-shaded area). (c) The largest negative frequency in unit of persistence time $\tau$ (red circles) as a function of persistence length $\ell$ is order $1$ at the crossover. Magenta symbols refer to lower packing fraction, i. e., $\phi=0.66$. Green squares are the structural relaxation time. The dashed blue line is $\tau |\omega^{-}_{max}|^2 = \ell$ (error bar indicates the statistical uncertainty over $100$ independent configurations).
}
\label{fig:dos}      
\end{figure*}

{Replacing $\ell$ in favor of $\tau$,} we thus stress that not only $T_{eff}(\lambda_\nu,\tau)$ makes sense for $\tau\to 0$, but also that the breaking of the concept of effective temperature is bounded to the structure of the energy landscape. For checking this fact, we have computed the spectrum of instantaneous and inherent normal modes. The corresponding density of states $\mathcal{D}_{inst,IS}(\omega)$ are shown in Fig. \ref{fig:dos}a. 
As a standard procedure, imaginary eigenmodes $\lambda^{(-)}$ have been mapped to negative frequencies, i. .e, $\omega^{(-)}=- \sqrt{ |\lambda^{(-)}| }$. At variance with instantaneous configurations, the frequency spectrum $\mathcal{D}_{IS}(\omega)$ of inherent configurations does not contain any negative frequency, meaning that it is linearly stable.
As $\ell$ increases, the average number of negative frequencies $\langle n^{(-)}\rangle $ increases (Fig. \ref{fig:dos}b). Both $\langle n^{(-)}\rangle $ and the (slowest) structural relaxation time $\tau_\psi$ approach a plateau value at the same crossover value $\ell_c \sim 1$. 

For connecting the crossover with instabilities in the energy landscape, 
we look at the behavior of the largest negative frequency $\omega^{(-)}_{max}$ as a function of $\ell$, as shown in Fig. \ref{fig:dos}c. The estimate from the effective temperature breaking is $ \tau_c |\lambda^{(-)}_{max}| \simeq 1$
as one can see,  we correctly obtain $\ell_c \sim 1$. 
We repeated the analysis at $\phi=0.66$ and the result fairly match the prediction (see Fig. \ref{fig:dos}c and Fig. \ref{fig:pd}b).
As in the case of equilibrium glasses, we can thus relate the features
of the energy landscape with  dynamical slowing down 
\cite{broderix2000energy,cavagna2001role,grigera2002geometric,castellani2005spin,angelani2000saddles}.
However, at variance with equilibrium, our picture suggests that the fluidization of glassy configuration in active system is connected to negative directions in the energy landscape rather than the marginal ones. 

\section*{Discussion}

Living systems as bacterial colonies or biological tissues are typically composed of self-propelled units of different sizes and different self-propulsive forces. In this work, we studied how this quenched disorder impacts the collective behavior of active particles.
As a model system, we have considered a  mixture of Active Brownian disks of different radii.
We have explored the phase diagram using as control parameters the persistence length $\ell$ of the active motion and the packing fraction
$\phi$. Similar to the case of monodisperse active systems, we have documented the presence of a MIPS coexistence region for  large persistence lengths. The phase separation extends from relatively small packing fractions ($\phi \sim 0.2$) to very high densities $\phi \sim 0.8$.

Geometrical frustration due to geometrical heterogeneities makes the system considered here a good model of glass \cite{PRX_Berthier}. We showed that this peculiarity did not get lost because of activity. In particular, we documented a glass transition at high density characterized by decoupling between the relaxation time of density fluctuations and that of the hexatic order parameter.

The study of the inherent structures has unveiled a deep connection between stationary configurations at small persistence lengths and the corresponding inherent ones. This connection turns out to be lost as $\ell$ increases. This is because MIPS, although it might be interpreted in terms of effective potential and equilibrium-like spinodal decomposition, it is an intrinsic non-equilibrium phenomenon caused by the self-propelled motion which cannot be completely reduced to an effective equilibrium description. 

The model introduced here is particularly suitable for studying the consequence of the competition between glassy dynamics, peculiar of equilibrium dynamics and known to play a pivotal role in dense active systems \cite{angelini2010cell,Angelini11}, and MIPS, a typical feature of active systems whose importance in bacterial colonies has been recently documented \cite{PhysRevLett.122.248102}. 

In this work, we have developed a deeper understanding of the concept of effective temperature in active matter, based on the analysis of the  spectrum of vibrations around its inherent structures. Using a simple picture based on the definition of effective temperature through an extension of the energy equipartition theorem \cite{Maggi14} that can be applied to dense active systems \cite{henkes2020dense,PhysRevE.84.040301,Bi2016}, we found that the breaking of the concept of effective temperature in active matter can be linked to the properties of the energy landscape of the instantaneous configurations. Since an instantaneous configuration is not optimal, the corresponding vibrational spectrum contains always negative eigenvalues. We observed that the magnitude of the largest of them can be used as a proxy, i. e., $\tau_c \sim 1/\lambda^{(-)}_{max}$, for estimating the crossover between \emph{active liquid}, from the point of view of the relaxation time, and a \emph{active glassy} regime, i. e., where the structural relaxation time starts increasing dramatically. The estimated value $\tau_c \sim 1$ is also in agreement with a heuristic argument that estimates the validity of the concept of $T_{eff}$ for vibrations smaller than the typical particles size $\tau \ll \langle \sigma \rangle / v_0 \sim 1$. 
Our approach, which is based on the properties of the vibrational energy landscape,  provides an interpretation of the mechanisms leading to the fluidization of frustrated glassy configurations. In particular, our picture suggests that, in spite the similarities between glassy dynamics in active and equilibrium-driven systems, the fluidization of the active glassy state is due to microscopic mechanisms different from those in equilibrium,  since the important parameter might be the magnitude of the largest negative eigenvalue rather than the number of quasi-stationary points \cite{cavagna1998stationary}. 
As a future direction, it might be interesting to study other aspects of the topological crossover in active glasses \cite{gradenigo2021non} and how it is connected with non-trivial spatial correlations of the velocity field \cite{PhysRevLett.124.078001,caprini2020hidden}. Moreover, while we have observed that the positional order in MIPS is strongly modified by the presence of polydisperse disks, less clear is how the phase boundaries of the MIPS region change with varying $f$.

Finally, the lack of any dynamical slowing down in the MIPS phase is consistent with the breaking of the concept of effective temperature. In particular, the MIPS regime lives always at high persistence lengths, and thus high activity, i. e., in a regime where non-equilibrium effects are important. This fact has strong implications, for instance, it follows that density and geometrical frustrations are not the only minimal ingredients for driving a dynamical arrest in Active Matter. More in general, MIPS seems to be incompatible with dynamically arrested phases. 
It seems crucial to understand whether or not the crossover from equilibrium-like glass and active regime might be captured by a Mode-Coupling Theory approach that suggests small vanishing values of the self-propulsion velocity of active tracers near the glass transition \cite{reichert2021transport}.

\section*{Methods}
We perform numerical simulations of a two dimensional Active Brownian particles composed of $N$ disks of different sizes, interacting via a short-range power-law potential, and confined on a square box of length $L$ with periodic boundary conditions.
\subsection*{Microscopic Model}
Due to their robustness against crystallization,
 polydisperse mixtures result to be good candidates as glass formers in a wide range of temperatures, even below the dynamical arrest temperature \cite{PRX_Berthier,wang2018low,khomenko2019depletion,PhysRevX.9.011049,PhysRevLett.122.255502,scalliet2019nature}. Here we consider a polydisperse mixture in two spatial dimensions where Active Brownian disks of different diameters $\sigma_i$ interact through a pair potential 
\BEA
v(r_{ij}) &=& \left( \frac{\sigma_{ij}}{r_{ij}} \right)^n + G(r_{ij}) \\
G(r_{ij})    &=& c_0 + c_2 \left( \frac{r_{ij}}{\sigma_{ij}} \right)^2 +  c_4 \left( \frac{r_{ij}}{\sigma_{ij}} \right)^4
\EEA
with $r_{ij} \equiv | \rr_i - \rr_j|$. We set the softness exponent to $n=12$. The coefficients $c_0$, $c_2$, and $c_4$ are chosen in a way that $v(r_c)=v^\prime(r_c)=v^{\prime\prime}(r_c)=0$,
where we have introduced the notation $v^\prime(r) = \frac{d v}{dr} $ and $v^{\prime\prime} (r) = \frac{d^2 v}{dr^2} $. 
For suppressing the tendency to demix, we consider non-additive diameters  
$
\sigma_{ij} = \frac{1}{2} \left( \sigma_i + \sigma_j \right) \left[1 - \epsilon | \sigma_i - \sigma_j |  \right] \;, \,
$
where $\epsilon$ tunes the degree of non-additivity \cite{zhang2015beyond,PRX_Berthier}. 
The cutoff is $r_c=1.25 \sigma_{ij}$, and $\epsilon=0.2$. The particle diameters $\sigma_i$ are drawn from a power law distribution $P(\sigma)$
with $\langle \sigma \rangle \equiv \int_{\sigma_{min}}^{\sigma_{max}} d\sigma \, P(\sigma) \sigma = 1$, with $\sigma_{min}=0.73$,
$\sigma_{max}=1.62$, and $P(\sigma)= A \sigma^{-3}$, with $A$ a normalization constant \cite{PRX_Berthier}.

The dynamical state of the $i-$th particle is given by its
position $\rr_i$ and by the orientation $\mathbf{e}_i$ of the self propelling force that, in two spatial dimension, is parametrized by the
angle $\theta_i$, i. e., $\mathbf{e}_i= \hat{\mathbf{x}} \cos \theta_i +\hat{\mathbf{y}} \sin \theta_i$, with $\hat{\mathbf{x}}$ and $\hat{\mathbf{y}}$ the unit vectors of the $x$ and $y$ axis, respectively.
The overdamped equations of motion for the $i-$th disk read
\begin{align}
    \dot{\rr}_i &=  v_0 \mathbf{e}_i + \mu \mathbf{F}_i + \boldsymbol{\zeta}_i\\
    \dot{\theta}_i &= \eta_i \; 
\end{align}
with $\langle \eta_i \rangle=0$ and $\langle \eta_i(t) \eta_j(s)\rangle=\frac{2}{\tau}\delta_{ij}\delta(t-s)$. 
The term $\boldsymbol{\zeta}_i$ is a thermal noise, i. e., $ \langle \zeta_i^\alpha \rangle = 0$,
and $\langle \zeta_i^\alpha(t) \zeta_j^\beta(s) \rangle = 2 \mu k_B T \delta_{ij} \delta^{\alpha \beta} \delta(t-s)$. 
The force $\mathbf{F}_i= \sum_{j \neq i} \mathbf{f}_{ij}$, with $\mathbf{f}_{ij}=-\nabla_{\mathbf{r}_i} v(r_{ij})$.
We perform numerical simulations with $k_B T = 0.01$, $\mu=1$, $v_0=1$. As control parameter we adopt the persistence length $\ell = \tau v_0 / \langle \sigma \rangle$ and the packing fraction $\phi = N A_s / A$, with $A_s = \pi \langle \sigma \rangle^2$.
The unit of length is fixed by $\langle \sigma \rangle = 1$, time is measured in units of $\tau$, and energy in units of $\epsilon=1$.
The system is enclosed in a square box of side 
$L=60 \langle \sigma \rangle$. 
For exploring the phase diagram, we have performed numerical simulations for $N \in [15^2,60^2]$ and persistence length $\ell \in [0.001,200]$. For evaluating the effect of polydisperisty on MIPS, we have performed numerical simulations of $N=80^2$ particles in a square box of side $L=120 \langle \sigma \rangle$ and we have changed the fraction of polydisperse disks $f \in [0,0.1]$. The definition of the observables is provided in SI.

\subsection*{Instantaneous and inherent configurations}
The inherent structures have been obtained minimizing the mechanical energy $\Phi= \frac{1}{2}\sum_{i \neq j} v(r_{ij})$.
Energy minimization is performed using the FIRE algorithm \cite{PhysRevLett.97.170201}. For gaining insight into the stability of inherent and instantaneous configurations, we have computed the normal modes by evaluating the $2 N$ eigenvalues of the Hessian matrix$\lambda_\kappa$, with $\kappa=1,..,2N$. The computations have been done using Python NumPy linear algebra functions \cite{mckinney2012python}.
The density of states of instantaneous and inherent configurations have been computed in a standard way, by introducing the eigenfrequency $\omega_\kappa=\pm \sqrt{| \lambda_\kappa |}$, with negative frequency corresponding to negative eigenvalues (that populate the spectrum only in the case of instantaneous configurations). The density of states reads $\mathcal{D}(\omega) = \mathcal{N}^{-1} \sum_\kappa \delta(\omega - \omega_\kappa)$, with $\mathcal{N}= 2 N - 2$.

\subsection*{Observables}
We indicate
with $\langle \mathcal{O} \rangle_s$ the average of the observable $\mathcal{O}$ over independent runs. 
We indicate with $\langle \mathcal{O} \rangle_t$ time-averaging in stationary states.
As order parameter describing the global and local properties of the system, we compute the packing fraction field $\phi(x,y,t)$ and the hexatic field $\psi_6(x,y,t)$.
$\phi(x,y,t)$ has been obtained discretizing the simulation box into a lattice of linear size $\zeta= 4 \langle \sigma \rangle$. We can thus define a probability distribution function $\mathcal{P}(\phi)$ of the local packing fraction. 
In the MIPS region homogeneous density profiles become unstable and thus $\mathcal{P}(\phi)$ develops a double-peaked structure.

The hexatic field at time $t$ can be computed starting from its microscopic definition and thus performing the Voronoi tesselation of the particle centers and defining the hexatic order parameter $\psi_6^i$ of the particle $i$ as
\begin{align}
    \psi_6^i(t) = \frac{1}{N_i} \sum_{k=1}^{N_i} e^{i 6 \theta_{ik}} \; ,
\end{align}
with $N_i$ the number of Voronoi neighbors to the cell $i$. The angle $\theta_{ik}$ is individuated by the two cell centers $i$ and $k$.
Again, we compute the distribution $\mathcal{P}(|\psi_6|)$ discretizing the simulation box into a lattice of linear size $\zeta$. 
We obtain additional information on the structural properties of the system measuring the pair distribution function
\begin{equation} \label{gofr}
    g(\mathbf{r}) = \frac{1}{N} \left\langle \sum_{i,j \neq i} \delta (\mathbf{r} - \mathbf{r}_j + \mathbf{r}_i)\right\rangle_{t,s} \; .
\end{equation}
and the $g_6(r)$ correlation function defined as
\begin{align}
    g_6(r) &= \langle \psi_6(\rr^\prime) \psi_6^*(\rr^\prime - \rr) \rangle_{s,t,\rr^\prime} \; .
\end{align}

As dynamical observables we measure the self-part of 
the Intermediate Scattering Function $F_{self}(q,t)$, 
and the time-correlation function of the hexatic order parameter $C_\psi(t)$ \cite{Lacevic2003,Berthier2011,flenner2015fundamental,massana2018active}.
The intermediate scattering function is
\begin{equation} \label{eq:fself}
    F_{self}(q,t) = \frac{1}{N} \left\langle \sum_i e^{ -i \qv \cdot (\rr_i (t) - \rr_i(0) ) } \right\rangle_s \; ,
\end{equation}
where the wave vector $\qv=(q_x,q_y)$ satisfies the periodic boundary conditions, i. e., $q_{x,y} = \frac{2 \pi}{L}(n_x,n_y)$, with $n_{x,y} = 0, \pm 1, \pm 2, ...$ (and avoiding the combination $n_x=n_y=0$).
The sample-to-sample fluctuations of $F_{self}(q,t)$ provides a measure of dynamical
heterogeneity through the four-point dynamical susceptibility $\chi_4(q,t)$.
The position of the peak $\chi_4(q,t)$ defines the typical time scale
 $t=\tau_4$ of dynamical heterogeneity. Map of displacements field has been computed as $\Delta \rr (x,y,\tau_4)$\cite{Berthier2011}. 
We also measure the relaxation time of shape fluctuations using $C_\psi(t)$ defined through the correlation function
\begin{equation} \label{eq:cpsi}
    C_\psi(t) = \frac{1}{C_\psi(0)} \left\langle \sum_i \psi_6^i(t) \psi_6^i(0) ^* \right\rangle_s \; .
    \end{equation}

To quantify the similarity between stationary and inherent configurations described through opportune coarse-grained variables $\mathbf{x}_{inst}$ and $\mathbf{x}_{IS}$ (for instance, $\mathbf{x}=|\psi_6|$), we have computed the Kullback–Leibler divergence between the probability distributions $\mathcal{P}_{inst}(\mathbf{x})$ and $\mathcal{P}_{IS}(\mathbf{x})$ that is defined as
\begin{align}
    D_{KL}[ \mathcal{P}_{inst} | \mathcal{P}_{IS} ] = \int d\mathbf{x} \,  \mathcal{P}_{inst}(\mathbf{x}) \log{ \frac{\mathcal{P}_{inst}(\mathbf{x})}{\mathcal{P}_{IS}(\mathbf{x})}} \; .
\end{align}

\subsection*{Data availability}
The data that support the findings of this study are available from the corresponding author on reasonable request.

\subsection*{Code availability}
The code is available from the corresponding author upon reasonable request.

\subsection*{Acknowledgments}
We thank L. Berthier for his critical reading of the manuscript.
M.P. has received funding from the European Union's Horizon 2020 research and innovation program under the MSCA grant agreement No 801370
and by the Secretary of Universities 
and Research of the Government of Catalonia through Beatriu de Pin\'os program Grant No. BP 00088 (2018). 
D.L. acknowledges MICINN/AEI/FEDER for financial support under grant agreement RTI2018-099032-J-I00.
I.P. acknowledges MICINN, DURSI and SNSF for financial support under
Projects No. PGC2018-098373-B-I00, No. 2017SGR-884,
and No. 200021-175719, respectively.

\subsection*{Author contributions}
M.P., D.L., and I. P. designed the research and discussed the results. M.P. performed simulations and data analysis. M.P, D.L., and I.P contributed to the writing of the manuscript.

\subsection*{Competing interests}
The authors declare no competing interests.

\bibliography{biblio}
\bibliographystyle{rsc}

\end{document}